\documentstyle[preprint,aps]{revtex} \tolerance=1000 \tighten

\begin{document}

\draft

\title{Electronic structure and  magnetic properties  of 
the linear chain cuprates Sr$_2$CuO$_3$ and Ca$_2$CuO$_3$ }

\author{H.\ Rosner, H.\ Eschrig, and R.\ Hayn}
\address{
Max-Planck-Arbeitsgruppe `Elektronensysteme', TU Dresden, D-01069
Dresden, Germany } 
\author{ S.-L. Drechsler \cite{dre} and  J.\ M\'alek \cite{mal}}
\address{ Institut f\"ur Festk\"orper- und Werkstofforschung Dresden e.V.,
 Postfach 270016, D-01171 Dresden, Germany }
\date{ \today             }

\maketitle

\abstract{

Sr$_2$CuO$_3$ and Ca$_2$CuO$_3$ are considered to be model systems of
strongly anisotropic, spin-$1/2$ Heisenberg antiferromagnets. We
report on the basis of a band-structure analysis within the local
density approximation and on the basis of available experimental data
a careful analysis of model parameters for extended Hubbard and
Heisenberg models. Both insulating compounds show half-filled nearly
one-dimensional antibonding bands within the LDA. That indicates the
importance of strong on-site correlation effects. The bonding bands of
Ca$_2$CuO$_3$ are shifted downwards by 0.7 eV compared with
Sr$_2$CuO$_3$, pointing to different Madelung fields and different
on-site energies within the standard $pd$-model. Both compounds differ
also significantly in the magnitude of the inter-chain dispersion
along the crystallographical {\bf a}-direction: $\approx$ 100 meV and
250 meV, respectively. Using the band-structure and experimental data
we parameterize a one-band extended Hubbard model for both materials
which can be further mapped onto an anisotropic Heisenberg model. From
the inter-chain dispersion we estimate a corresponding inter-chain
exchange constant $J_{\perp} \approx$ 0.8 and 3.6 meV for Sr$_2$CuO$_3$ and
Ca$_2$CuO$_3$, respectively. Comparing several approaches to
anisotropic Heisenberg problems, namely the random phase spin wave
approximation and modern versions of coupled quantum spin chains
approaches, we observe the advantage of the latter in the reproduction
of reasonable values for the N\'eel temperature $T_N$ and the
magnetization $m_0$ at zero temperature. Our estimate of $J_{\perp}$
gives the right order of magnitude and the correct tendency going from
Sr$_2$CuO$_3$ to Ca$_2$CuO$_3$. In a comparative study we also include
CuGeO$_3$.  }


\section{Introduction}

\newcommand{\sco}{Sr$_2$CuO$_3$}
\newcommand{\cco}{Ca$_2$CuO$_3$}
\newcommand{\gco}{CuGeO$_3$}
\newcommand{\mbohr}{$\mu_{\rm\small Bohr}$}
\newcommand{\etal}{{\em et al.\ }}

Initiated by the discovery of high-$T_c$ superconductivity in cuprate
compounds, there is a renewed and growing interest in the electronic
and magnetic properties of quasi one-dimensional (1D) structures near
half band filling. Anionic quasi 1D CuO$_3$ chains of oxygen corner
sharing CuO$_4$ plaquetts are present in \sco\ and in \cco\
\cite{TM,HH}, and CuO$_2$ chains of edge sharing plaquetts are present
in \gco\ \cite{hase93}. Moreover, the first two of these compounds can
be regarded as `parent' compounds of more complex structures as the
double chain compound SrCuO$_2$ (which, in notation Sr$_2$Cu$_2$O$_4$,
is obtained from \sco\ by replacing the CuO$_3$ chain with a
Cu$_2$O$_4$ double chain), and of a whole family of celebrated
multi-leg ladder compounds Sr$_{2n}$Cu$_{2n+2}$O$_{4n+2}$
\cite{ishida96}. A detailed theoretical description of the `parent'
compounds is naturally a prerequisite for understanding all those
materials. It is also desirable in view of the related two-dimensional
(2D) cuprate structures of the high-$T_c$ materials
\cite{tokura90,yoshida91,keren93,ami95,eggert96,motoyama96,misochko96},
and particularly in view of possible dynamical stripe superstructures
in the latter \cite{emery96}.

Experimentally, \sco\ and \cco\ are found to be the best realizations
of the quasi 1D spin-1/2 antiferromagnetic Heisenberg model
(AHM). Their N\'eel temperatures, $T_N \approx $ 5K for \sco\ and $T_N
\approx $ 9K for \cco, are very low compared to the intra-chain
exchange integrals $J_{\|}\sim $ 0.2 eV, and the ordered moments
($<$0.1 \mbohr) are extremely small
\cite{kojima96,yamada95,kojima97}. The value $J_{\|}$ = 190 meV for
\sco\ \cite{motoyama96} appears to be the record value of an exchange
integral among all known quasi 1D antiferromagnets. The correct
description of the physics of a magnetic quasi 1D system with a weak
magnetic inter-chain interaction has recently attracted much
theoretical attention, see, e.g.,
Refs.\ \cite{schulz96,affleck94,affleck96,wang96}. For all these reasons,
\sco\ has been announced to become a `superstar' in the field of
low-dimensional magnetism in near future \cite{tsvelik95}.

In the work of Ami \etal \cite{ami95}, with the assumption of only a very weak
dipolar inter-chain interaction constant $J_\perp \approx$ 0.01 meV and
on the basis of RPA spin wave theory, for \sco\ a N\'eel temperature
below 0.03K \cite{remark4} was conjectured. At that time, due to resolution problems
and because of the very small ordered moment $<$0.1\mbohr, neutron
diffraction on powders failed to detect antiferromagnetic ordering
down to 1.5K. However, the relatively small inter-chain distances of
3.3 to 3.5\AA\ suggest that direct inter-chain hopping, which leads to
a much stronger kinematic exchange interaction, cannot be
neglected. The discussion of consequences for the magnetic properties
forms a main issue of the present paper.

While numerous band-structure calculations \cite{Pickett} and
tight-binding parameterizations of one- and multi-band Hubbard model
Hamiltonians \cite{Hybertsen} for the quasi 2D cuprate structures can
be found in the literature, we are aware of only two band-structure
calculations for a quasi 1D cuprate structure, both for \gco\
\cite{mattheiss94,popovic95}. On the basis of a wealth of available
experimental data for \gco\ large efforts are currently directed to
the parametrisation of phenomenological antiferromagnetic spin-1/2
Heisenberg models and extensions including frustration in the
next-nearest neighbor intra-chain exchange
\cite{riera95,regnault96,kuroe96,gros96}. To our knowledge, no
parameterization on the level of Hubbard-type models has been
undertaken as yet, although estimates within the Anderson impurity
model \cite{parmigiani96} were pointing to strong correlation. The
understanding of CuO$_3$ chain substances is less developed,
especially with respect to inter-chain interactions, and we shall
present here a comparative analysis of both cases.

In contrast to the antiferromagnetic ordering of \sco, the less
anisotropic compound \gco\ exhibits a spin gap state below
$T_{SP}$=14.2K which is accompanied by the occurrence of a period-2
superstructure with a very small dimerization amplitude of
$u_0\approx$ 0.007\AA. Therefore it has been interpreted as a
spin-Peierls (SP) state. The SP state is supported by frustrated
second neighbor exchange \cite{riera95}. When doped with Zn for Cu or
with Si for Ge, also a coexisting N\'eel state has been found below
4.5K \cite{hase96,poirier95}. The thereby observed magnetic moment,
for example of 0.23\mbohr\ in Cu$_{1-x}$Zn$_x$GeO$_3$, is
significantly larger than the corresponding value of 0.06\mbohr\
observed in \sco\ \cite{kojima96}, suggesting a much larger anisotropy of the
latter compound.

Within the frame of strongly anisotropic three-dimensional Heisenberg
models, all considered compounds should be described by a dominating
intra-chain exchange coupling, a small exchange coupling in direction
of the shortest inter-chain spacing which reduces the strong quantum
fluctuations and provides a non-zero ordered magnetic moment in the
ground-state, and a generally very small inter-chain coupling in the
third direction to ensure a non-zero N\'eel temperature in accord with
the Mermin-Wagner theorem. By applying the results of available
theoretical approaches to the anisotropic two- and three-dimensional
Heisenberg models \cite{schulz96,affleck94,affleck96,castro96} one may
extract phenomenological estimates of the exchange parameters for the
chain cuprates, which then can be compared to results of electronic
structure theory. Interestingly, different approaches yield
significantly different predictions. The accurate determination of the
N\'eel temperature of a strongly anisotropic Heisenberg magnet is
still an unsolved and challenging theoretical task. We review here two
frequently used approaches and compare their predictions from our
estimates of exchange parameters with experimental data.

In the following Sec.\ II we present band-structure results for \sco\
and \cco\, which were obtained by applying a
linear-combination-of-atomic-orbitals scheme to self-consistently
solve the Kohn-Sham equation with the local-density approximation to
the exchange and correlation potential (LDA-LCAO). This approach
provides us in a most natural way with tight-binding parameters. Using
these results and experimental estimates for short-range correlation
effects, in Sec.\ III a single CuO$_3$ chain is represented by an
extended Hubbard model and compared to the situation with \gco. The
magnitude of inter-chain exchange is estimated on this basis. In Sec.\
IV, estimates of N\'eel temperatures and ordered magnetic moments are
derived by applying both standard RPA spin wave theory and modern
quantum spin chain theory. The results are summarized in Sec.\ V, and
an outlook is given.

\section{Band-structure and inter-chain transfer}

The crystal structure of the isostructural compounds \sco\ and \cco\
is depicted in Fig.\ 1. Chains of oxygen corner sharing CuO$_4$
plaquettes run along the {\bf b}-direction. The in-chain Cu-O
bond-length is 1.96 \AA\ and practically the same in both compounds
while the Cu-O bond-lengths in {\bf c}-direction differ: 1.95 \AA\ in
\sco\ and 1.89 \AA\ in \cco. The two inequivalent oxygen sites are
referred to as chain oxygen and side oxygen in the following. The
shortest inter-chain distance occurs in {\bf a}-direction and differs
substantially for both cases: 3.49 \AA\ in \sco\ and 3.28 \AA\ in
\cco.

The self-consistent LDA-LCAO method has been applied to both compounds
with a minimum basis treating the Cu-($4s,4p,3d$), O-($2s,2p$),
Sr-($5s,5p,4d$) and Ca-($4s,4p,3d$) orbitals as local valence basis
states and the lower orbitals as core states. The crystal potential is
calculated from overlapping spherical site densities. All basis states
are calculated in the spherical site contribution to the crystal
potential and recalculated in each iteration step. The valence basis
orbitals have been compressed by an additional attractive potential to
reduce the overlap among them \cite{Eschrig}. Due to the relatively
open crystal structure two empty spheres per unit cell have been
introduced with empty sphere $s$ and $p$ orbitals at each site. For the
exchange and correlation potential the parameterization of von Barth
and Hedin was chosen and it has been calculated in atomic sphere
approximation. We show in Fig.\ 2 the band-structure and in Fig.\ 3 the
density of states (DOS) of \sco. The corresponding results for \cco\
are similar. The quantitative differences between both compounds are
discussed below. To check the LCAO-band-structure by another method, we
performed also calculations using the linear muffin-tin orbital (LMTO)
approximation. We found no substantial differences, only the overall
bandwidth of the whole $pd$ band complex was found to be somewhat
smaller in the LMTO results (see numbers below).

As expected from simple chemical considerations of covalency, there is
a single, well separated, nearly one dimensional, half-filled
antibonding band crossing the Fermi level with large dispersion in
{\bf b}-direction (see Fig.\ 2).  The width of this band is about
$2.2$ eV (LCAO) or $2.0$ eV (LMTO) for both compounds.  The
characteristic quasi-1D van Hove singularities near the band edges are
clearly seen in the DOS (see Fig.\ 3). A tight-binding analysis of the
orbitals involved shows that in first approximation the Cu
$3d_{z^2-y^2}$ as well as the side oxygen $2p_z$ and the chain oxygen
$2p_y$ orbitals are of direct relevance. Only a negligible admixture
of Cu 4$s$ states can be detected near both edges of this antibonding
half-filled band. Its weight as determined by the ratio of the
corresponding areas under the DOS curves (see insert of Fig.\ 3) is
relatively small (0.3 per cent for \sco), but it increases to a weight
of 2 per cent for the Ca compound.

The metallic behaviour of the LDA band-structure is in sharp contrast
to the experimental observation of large optical gaps $\sim$ 2 eV
which are comparable to the bandwidth obtained above. This points to
the necessity of dealing explicitly with the strong on-site Coulomb
repulsion at the copper-site. The experimental gap cannot be explained
by a spin density wave since it is large and persists also above the
N\'eel temperature $T_N$. Instead we have to anticipate the situation
of a charge transfer gap between valence states of mostly oxygen
character and a copper upper Hubbard band above the Fermi level.

In analogy to cuprates with CuO$_2$ planes the construction of a
multi-band, Hubbard-like model Hamiltonian would therefore be
desirable. However, it is well known that such a Hamiltonian can be
projected to an effective one-band picture which properly describes
the low-energy physics \cite{fedro92,drechsler96}. The existence of a
well isolated, one dimensional band in the present situation (shown in
Fig.\ 4 in more detail) suggests such a possibility all the more.  We
assume that the parameters for the one band description can be
determined by fitting the band of Fig.\ 4 to a dispersion of the form
\begin{equation}
\varepsilon (\vec{k}) = -2t_{1,LDA} \cos (k_yb) -2t_{2,LDA} \cos (2k_yb)
-2t_{\perp}\cos k_xa  \\ 
\label{dispers}
\end{equation}  
which yields the values listed in the Table \cite{remark1}.

In Fig.\ 4 the dispersion in the {\bf a}-direction is clearly visible
as an energy increase with increasing $k_x$ by nearly the same amount
both at the bottom and the top of the band. This dispersion is present
through the band and gives a value of $t_{\perp}=25$ meV. 
To be more accurate we determined $t_{\perp}$ from the dispersion at
the Fermi level which is shown in the insert of Fig.\ 4.
The corresponding dispersion for \cco\ is significantly larger, by a
factor of 2.5 (2.0 in LMTO). The smaller lattice constant of \cco\
leads to an increase of the inter-chain overlap of the Cu-$d$ and O-$p$ basis
orbitals, but this effect alone is too small to explain the strong
enhancement. 
We have checked that the transfer in {\bf a}-direction goes dominantly
via the cation Sr and Ca, respectively.
The two-center Hamilton matrix elements between side oxygen and Ca are
two times larger than the corresponding ones for Sr.

Dispersion in the {\bf c}-direction is found within the level of
accuracy of the band-structure calculations only ($\sim 5$ meV). This
practically missing dispersion in the {\bf c}-direction indicates also
that the inter-chain hopping in {\bf a}-direction takes place
horizontally only, with no zigzag contribution in (111) direction from
side oxygen to side oxygen.

The comparison of the band structures of the Sr and the Ca compounds
shows yet another interesting feature: namely, the bonding bands 
between -8.5 eV and -11.5 eV for \sco\ are
shifted downwards by 0.6 to 0.7 eV for the Ca compound indicating a
significantly different Madelung field. For the same reason the side
oxygen in the Ca-compound comes out more negatively charged although
the hybridization of its $p_z$-orbital with the Cu $3d_{z^2-y^2}$
orbital is slightly increased due to the shortened bond-length. This
different on-site energies may have an influence on the different
charge transfer energies as discussed in the next section.

We also analyzed in the same manner (Eq.\ \ref{dispers}) the
linearized augmented plane waves (LAPW) energy bands for \gco\
reported by Mattheiss \cite{mattheiss94}. \gco\ differs from the chain
cuprates considered above in the structure of the chains. The CuO$_2$
chains of oxygen edge-sharing CuO$_4$ plaquetts of \gco\ result in a
more complex highest antibonding band in which two O-$p$ states per
oxygen site hybridize with the Cu-$d$ orbital. The more complex
composition of the crystal also manifests in the band-structure.  In
particular, due to the sizeable inter-chain interaction mediated by Ge
and due to the presence of two chains per unit cell there are two
differently filled split-off antibonding bands. For our qualitative
comparison with the above considered chains (CuO$_3$ chain), we
replace them by one half-filled band for the sake of simplicity.  The
tight-binding parameters of CuGeO$_3$ (Table) contain a significantly
smaller nearest neighbor transfer integral $t_{1,LDA} \approx 0.25$ eV
and an anomalously large next nearest neighbor integral \cite{remark2}
$t_{2,LDA} \approx 0.67 t_1 = 0.18$ eV estimated from Fig.\ 2 of Ref.\
\cite{mattheiss94}.  The large difference of the transfer integrals
$t_{1,LDA}$ between chains of corner sharing plaquetts and \gco\ should be
related to the efficient 180$^o$ Cu-O-Cu hopping for the former
($\sigma $ $p-d$ bond) compared with the inefficiency of non$\sigma $
$p-d$ hopping (nearly 135$^o$ (45$^o$)) for \gco; for further
details see Ref.\ \cite{khomskii96}. This special structure explains
also the relative large next nearest neighbor transfer integrals $t_2$
in \gco\ due to the involved effective $\sigma$ $p_z$-$p_z$
hopping along the chain. The inter-chain hopping $t_{\perp}$ can be
deduced from the dispersion in the {\bf b}-direction of the LAPW
energy bands of CuGeO$_3$ and is of the same order as in Sr$_2$CuO$_3$
and Ca$_2$CuO$_3$.

Based on both the available experimental data and the band
structure information obtained here, a semi-microscopic strong
correlation model will be constructed which then can be mapped
approximately onto a spin-1/2 Hamiltonian to describe the magnetic
properties. This is the objective of the next section.

\section{Microscopic description in terms of the extended Hubbard and
anisotropic Heisenberg models}

\subsection{General relations}

Here we parameterize the well-known extended Hubbard model for
one single chain with hopping terms to first ($t_1$) and second
neighbors ($t_2$):
\begin{eqnarray}
H&=&-\sum_{m;j=1,2;s} t_j \left( c^{\dagger}_{m,s}c_{m+j,s} +h.c. \right)
+\frac{U}{2} \sum_{m;s} c^{\dagger}_{m,s}c_{m,s}c^{\dagger}_{m,-s}c_{m,-s}
\nonumber \\
& &+\sum_{m;j=1,2} V_jn_mn_{m+j} 
-\mid K \mid \sum_i\vec{S}_i\vec{S}_{i+1}, 
\label{hubbard}
\end{eqnarray}
where $n_m =\sum_sc_{m,s}^{\dagger}c_{m,s}$ is the density operator
and $s$ denotes the spin index. We included in (\ref{hubbard}) a
small, but non-negligible direct ferromagnetic exchange which
naturally appears if we map a multi-band, Hubbard-like Hamiltonian to a
one band model \cite{fedro92}. Its necessity and its main effects will
be discussed below.

For the low-energy physics, at half-filling the extended Hubbard model
(\ref{hubbard}) can be replaced to leading order in $t/U$ by a
spin-1/2 Heisenberg chain. It includes also a second neighbor exchange
$J_2$ \cite{schulten76} and reads
\begin{eqnarray}
&&H=J_1 \sum_{i}\vec{S_i}\vec{S}_{i+1} +
J_2 \sum_i \vec{S_i}\vec{S}_{i+2} \; , \nonumber \\
&& 
J^{AF}_1= \frac{4t_1^2}{U-V_1} \; , \qquad
J_2 = \frac{4t_2^2}{U-V_2} \; ,
\label{longheis}
\end{eqnarray}
where the effective exchange integral $J_1$ of the spin-1/2 Heisenberg
Hamiltonian of cuprates is reduced from the predominant
antiferromagnetic superexchange part by the ferromagnetic contribution
(\ref{hubbard})
\begin{equation}
J_1\approx -\mid K \mid   +J^{AF}_1.
\label{totex}
\end{equation}
Notice that within this approach $J_2$ yields a competitional
(frustrated) character to the usually dominant short range
antiferromagnetic correlations which are established by $J_1$.  That
term is especially important for CuGeO$_3$.

The two main parameters $U$ and $t_1$ of the effective extended
Hubbard model are directly related to the optical gap $E_g$ and the
exchange integral between nearest neighbors $J_1$ which are
experimentally accessible. The following analysis is considerably
simplified if the materials of interest are in the strongly correlated
limit $U>4t$ and excitonic effects at zero momentum transfer $q$ are
not very strong i.e.\ $U \gg t_j >\ V_j ;\ j=1,2$. The parameter sets
derived below support such a point of view.  We take into account the
effect of the intersite Coulomb interaction $V_1$ by renormalizing the
on-site correlation in the form $U_{eff}=U-V_1$.  Then we may use the
optical gap $E_g$ obtained from the Bethe-Ansatz solution for the pure
1D Hubbard model given by Ovchinnikov \cite{ovchinnikov69},
\begin{eqnarray}
E_g&=&\frac{16t_1^2}{U_{eff}}\int\limits_1^{\infty}
\frac{\sqrt{x^2-1}dx}{\sinh(2\pi t_1x/U_{eff})}  \nonumber\\
&\approx &U_{eff}-4t_1+  2 \mbox{ln} 2 J^{AF}_1\quad \mbox{for}\quad U_{eff}
\gg t_1;
\quad  J^{AF}_1=4t_1^2/U_{eff}\nonumber.\\
\label{ovch}
\end{eqnarray}
In the strong coupling case Eq.\ (\ref{ovch}) can be transformed to
the useful relation
\begin{equation}
t_1=0.5J^{AF}_1\left(1+\sqrt{E_g/J^{AF}_1 +1- 2 \mbox{ln} 2}\right).
\label{tj}
\end{equation}
It has been assumed that the smaller parameters $t_2$ and $\mid K
\mid$, i.e.\ the hopping to second neighbors and the ferromagnetic
exchange in (\ref{hubbard}), have no substantial influence on the
charge transfer gap (but $t_2$ enhances the spin gap in the
spin-Peierls state).

The presence of a weak second neighbor exchange can be approximately
described in some cases by an effective renormalized nearest neighbor
exchange integral \cite{gottlieb91,muellerfledder96}
\begin{equation}
J=J_1-rJ_2 \approx J_1-J_2,
\label{frust}
\end{equation}
where $r=1$ according to Ref.\ \cite{gottlieb91} and $r=1.12$
according to Ref.\ \cite{muellerfledder96}.  Recently, Stephan and
Penc \cite{stephan96} predicted a strong narrow excitonic peak in the
density-density response function $N(q,\omega)$ of the EHM in the
strong coupling limit at the zone boundary $q=\pi /b$:
\begin{equation}
 \omega_{ex}(\pi/b) = U-V_1, 
\label{stephan}
\end{equation}
provided $V_1 > 0$. 

\subsection{Parameter assignment}

In principle, we can determine $U_{eff}$ and $t_1$ from the
experimentally measured $E_g$ and $J$ values using Eqs.\
(\ref{totex}--\ref{frust}) which are presented graphically in Fig.\ 5. In
the case of Sr$_2$CuO$_3$, very recently also the narrow excitonic
peak at the zone-boundary (\ref{stephan}) and with it $U_{eff}$ were
determined experimentally \cite{neudert97}.

However, to the best of our knowledge, the available experimental
information on all three systems is incomplete or contradicting each
other. For instance, for \cco\ the charge gap determined from the
maximum of $\Im m \varepsilon (\omega)$ is $E_g=2.1$
eV\cite{tokura90}, but a direct measurement of the $J$ value from
the magnetic susceptibility does not exist. Interpreting the
midinfrared absorption as a phonon-assisted two magnon process a value
of $J=255$ meV was reported \cite{suzuura96}. For \sco\ the
experimental $J$ values range from 140 to 260 meV
\cite{eggert96,motoyama96,suzuura96}. In the following we shall use
190 meV as a representative value. According to quite recent data for
this system the one-dimensional charge transfer gap $E_g$(Sr)$\approx
1.9 \pm 0.1$ \cite{neudert97} might be somewhat smaller as compared to
the Ca compound. Strictly speaking, the optical absorption sets in
already near 1.5 eV \cite{kotani96,sarma96,neudert97}. The elucidation
of the observed broadening of the expected 1D van-Hove singularity in
terms of the inter-chain interaction, quantum fluctuations , disorder,
and/or excitonic and other many-body effects is a difficult problem
beyond the scope of the present paper.

Taking this situation into account, we use the available experimental
data and also our band-structure results to derive a consistent
parameter set of (\ref{hubbard}) ($t_1$, $t_2$, $U_{eff}$, $V_1$ and
$\mid K \mid$) for each of the three substances, separately. Vice
versa, the demand of internal consistency weights the experimental
information.

\subsubsection{S\lowercase{r}$_2$C\lowercase{u}O$_3$}

Recent electron energy loss spectroscopy data of Neudert {\it et al.}
\cite{neudert97} allow to determine $U_{eff}=3.15 \pm 0.1 $ eV from
the maximum of $\Im m\varepsilon(\omega)$ at the zone boundary
(\ref{stephan}). At the same time $E_g$ was measured to be 1.9 $\pm
0.1$ eV from the data at small momentum. Similarly $E_g=1.92$ eV was
found from the Raman resonance energy \cite{misochko96} observed for
diagonal in-chain (yy) polarization only. Our aim is to derive values
for the magnetic coupling constants from Eqs.\
(\ref{totex}--\ref{frust}) with the use of experimental values of
$U_{eff}$ and $E_g$. Since it turns out that the derived $J$-values
depend sensitively on $U_{eff}$ and $E_g$, we consider two sort of
extreme cases. From Eq.\ (\ref{ovch}) we obtain $t_1=0.410 $eV with
$U_{eff}=$3.15 eV and $E_g=1.8$ eV (lower bound). That corresponds to
$J^{AF}_1$=213 meV. According to Eqs.\ (\ref{totex},\ref{frust}) that
value has to be reduced by the frustrated next nearest neighbor
exchange $J_2$ of about 12 meV (corresponding to $t_2=100$ meV from
our tight binding fit) and by the ferromagnetic contribution $\mid K
\mid$ before it can be compared with the total experimental exchange
integral $J=190 \pm 17$ meV \cite{motoyama96}. Thus we can estimate a
direct ferromagnetic exchange of $\mid K \mid \approx 11 \pm 17$ meV.
The slightly smaller $K$ value compared with 35 meV for La$_2$CuO$_4$
obtained in Ref.\ \cite{fedro92} might be attributed to the shorter
Cu-O bond-length of 1.89 \AA\ for the latter compound. The so derived
parameter set is listed in the Table. We derived a second parameter
set with a considerable smaller value of $J_1$ by taking $E_g=1.95$ eV
and $U_{eff}=3.25$ eV near the upper bounds of the experimental
results.  We obtain from (\ref{ovch}) $t_1=0.394$ eV and
correspondingly $J^{AF}_1=190$ meV. Such a parameterization is
compatible with the total exchange integral 147$^{+13}_{-9}$ meV
\cite{eggert96} derived from the magnetic susceptibility data.  We
note that both parameterizations are incompatible with the large $J$
values of 246 meV \cite{lorenzana97} and 261 meV \cite{suzuura96}
derived from midinfrared optical absorption data (the small
differences between the latter values arise mainly from the adopted
phonon frequency of 70 and 80 meV, respectively, involved in the
phonon-assisted absorption process) provided there is no sizeable {\it
ferromagnetic} second neighbor exchange over-compensating the
ferromagnetic nearest neighbor contribution $\mid K \mid $ and the
antiferromagnetic next nearest neighbor superexchange
$J_2=4t_2^2/U_{eff}$. Anyhow, the elucidation of the microscopic
origin of the apparent discrepancy between the magnetic susceptibility
and the midinfrared optical absorption data analyzed in terms of the simple
nearest neighbor spin-1/2 Heisenberg model remains a challenging
problem.

\subsubsection{C\lowercase{a}$_2$C\lowercase{u}O$_3$}

The slightly larger charge transfer gap of 2.1 eV suggests also an
enhanced $U_{eff}$-value in comparison with the Sr-compound. That
means that it is again difficult to find a reasonable parameterization
which is compatible with the large $J$ value of 254 meV from
midinfrared absorption data. Due to the lack of experimental
information on the magnetic susceptibility we use in the following our
theoretical estimate of 160 meV for the $J$ value of Ca$_2$CuO$_3$
\cite{drechsler96}. Adding a ferromagnetic contribution of $\mid K
\mid \approx 30$ meV (of the same order as for La$_2$CuO$_4$) and a
frustrating $J_2 \approx 10$ meV we may find $J_1^{AF}=200$ meV. Of
course, in the given case this should be considered as a very rough
estimate. Then, together with $E_g=2.1$ eV, we calculate from
(\ref{tj}) $t_1=418$ meV. According to (\ref{ovch}) that corresponds
to $U_{eff}=3.5$ eV, showing the expected enhancement. On the level of
the $pd$-model the reason for the enhanced effective on-site
interaction should be traced back to a larger $\Delta \varepsilon_{pd}
\approx U_{eff}$.  It seems to be related to a Madelung effect caused
by the difference in the lattice parameters of the Sr and Ca compound,
respectively.  This point of view is corroborated by our band-structure
calculations discussed in the previous section: for Ca$_2$CuO$_3$ the
distance between the half-filled antibonding band and the filled bonding
bands is larger by 0.7 eV compared to the Sr case.

\subsubsection{Inter-site Coulomb repulsion and comparison with C\lowercase{u}G\lowercase{e}O$_3$}

According to the microscopic calculations of Geertsma and Khomski
\cite{khomskii96} the total nearest neighbor exchange integral of
CuGeO$_3$ $J_1$(Ge)=11.6 meV can be decomposed into an
antiferromagnetic contribution of $J_1^{AF}=30.4$ meV and into a
relatively large ferromagnetic one of $\mid K \mid=18.8$ meV. Let us
stress again that the effective nearest neighbor transfer integral
$t_1$ and the on-site interaction $U_{eff}$ are directly related to
the antiferromagnetic part only. For CuGeO$_3$ charge transfer gaps
of 3.66 eV \cite{parmigiani96} and 1.25 eV \cite{terasaki95} have been
reported \cite{remark2}. Using $E_g=3.7$ eV and $J^{AF}_1=30.4$ meV,
the main parameters of the extended Hubbard model come out as
$t_1=0.187$ eV and $U_{eff}=4.34$ eV. Interestingly, the latter value
nearly perfectly coincides with the charge transfer energy
$\Delta$=4.2 eV found out from the XPS data analyzed within the
Anderson impurity model in Ref.\ \cite{parmigiani96}.  Within a
$pd$-model the significantly enhanced corresponding $\Delta
\varepsilon_{pd}$ value should be attributed to the Ge-cations located
near the CuO$_2$ chain oxygens. This point of view is supported by the
following observations. In compounds like Sr$_{14}$Cu$_{24}$O$_{41}$
where the CuO$_2$ chains under considerations are surrounded by earth
alkaline cations the corresponding charge gap is reduced to about 2.8
eV \cite{osafune97}. For that compound, $\Delta\varepsilon_{pd}$ as
calculated within the ionic point charge model amounts to 3.7eV
\cite{mizuno96}.

Comparing the data collected in the Table we suggest that
Ca$_2$CuO$_3$ should be somewhat stronger correlated than its Sr
counterpart.  Without doubt, the most strongly correlated compound
among them all is CuGeO$_3$ having the smallest transfer integral
$t_1$(Ge)=0.187 eV and the largest $U_{eff}$(Ge) = 4.34 eV. The large
ratios $U_{eff}/t_1$ obtained in all three cases (7.7(Sr), 8.4(Ca),
and 23(Ge)) justify {\it a posteriori} the use of Eq.\ (\ref{ovch}).

The difference between $t_1$ and $t_{1,LDA}$ may be explained by a
renormalization of the transfer integral $t_1$ by the inter-site
Coulomb interaction $V_1$. Within the Hartree-Fock approximation, the
correction due to $V_1$ leads to a renormalized effective hopping
integral $t_1+p V_1$ with the bond order $p\sim 2/\pi$. This
renormalized hopping integral can be compared with $t_{1,LDA}$ where
the inter-site Coulomb interaction is already partially taken into
account. From $t_{1,LDA}=t_1+pV_1$ and the data of $t_1$ and
$t_{1,LDA}$ given in the Table we may determine $V_1=0.21$ eV for
Sr$_2$CuO$_3$, $V_1=0.16$ eV for Ca$_2$CuO$_3$, and $V_1$=0.1 eV for
CuGeO$_3$ (here $t_1=0.187$ eV as estimated above has been adopted).
Thus, the inter-site Coulomb interactions $V_1\stackrel{<}{\sim}$ 0.2
eV of all three compounds are quite close to each other and fulfill
the relation $V_1 \ll U$. Notice that these numbers for $V_1$ roughly
agree with the corresponding 2D-values $0.11$ eV or $0.17$ eV given in
Refs.\ \cite{fedro92,feiner96}, respectively, and the estimate based on
the four-band model for CuO$_3$ chains \cite{drechsler97}: $V_1
\approx n_dn_pV_{pd} + n_d^2(V_{dd}-V_{2,dd}) \approx 0.23$ eV, where
typical occupation numbers $n_d\approx 0.7, n_p\approx 0.13$ and
$V_{pd}=1.2 $eV for the copper oxygen inter-site Coulomb interaction
and $V_{1,dd}=0.5$ ($V_{2,dd}=0.25$) eV for the (next) nearest
neighbor copper copper inter-site Coulomb interaction have been taken.

The value for $t_2$ in the Table was either taken from the fit to the
band-structure data ($=t_{2,LDA}$ for Sr$_2$CuO$_3$ and Ca$_2$CuO$_3$ ) or inferred
from the experimentally known value for $J_2=4.3$ meV \cite{riera95}
for CuGeO$_3$ using Eqs.\ (5).

\subsubsection{Inter-chain exchange}

Now, we would like to give a first estimate of the magnetic couplings
between chains. The inter-chain exchange interaction $J_{\perp}$ in the
{\bf a}-direction for Sr$_2$CuO$_3$ and Ca$_2$CuO$_3$ (corresponding to the {\bf
b}-direction in CuGeO$_3$) will be approximated by
\begin{equation}
J_{\perp}=\frac{4t_{\perp}^2}{U_{eff}},
\label{interchain}
\end{equation}
where we assumed for simplicity the same inter-site Coulomb interaction
$V_1$ within the chain and perpendicular to it. The corresponding
values are listed in the Table. The discussion above about a possible
direct ferromagnetic exchange which leads to a systematic reduction of
exchange integrals suggests that these values should be considered as
upper bounds. In the case of CuGeO$_3$ the so determined $J_{\perp}=1$
meV can be compared with experimental data from neutron scattering
\cite{nishi95,regnault96} $J_{\perp} \approx 1.1$ meV showing a
reasonable agreement.  Notice that in the case of Sr$_2$CuO$_3$ our
inter-chain interaction exceeds the dipolar interaction evaluated in
Ref.\ \cite{ami95} by two orders of magnitude.

The magnitude of the weakest interaction $J_{\perp,c}$ in the {\bf
c}-direction is difficult to estimate theoretically for several
reasons. It was already discussed that a band-structure hopping
integral cannot be given at present. In any case it may be expected
that $J_{\perp,c}$ is smaller than the other exchange integrals by
several orders of magnitude and is difficult to calculate in any
case. Treating $J_{\perp,c}$ therefore as a purely phenomenological
parameter in the following, we will use to be specific (if necessary)
the same value for Ca$_2$CuO$_3$ as has been evaluated in Ref.\
\cite{ami95} for Sr$_2$CuO$_3$, adopting the dipolar interaction for
$J_{\perp,c} \approx 10^{-4}$ meV.

\section{Some aspects of the N\'eel state }

The magnetic properties of undoped cuprates (i.e. one hole per Cu site
in the standard $pd$-model) are usually described by the anisotropic
spin-1/2 antiferromagnetic Heisenberg model
\begin{equation}
H=\sum_{<i,j>}J_{ij} \mbox{\bf S}_i \mbox{\bf S}_j \quad , 
\label{heisi}
\end{equation}
with $J_{ij}=J_{\parallel} ( = J_1)$ for $(ij)$ beeing nearest
neighbors in the chain direction (that is the {\bf b}-direction for
Sr$_2$CuO$_3$ and Ca$_2$CuO$_3$) and $J_{\perp}$ for nearest neighbor
copper sites in {\bf a}-direction (see Fig.\ 1). The weakest
interaction will be denoted here by $J_{\perp,c}$. According to the
results of the previous sections and from the experimental data we
know that Sr$_2$CuO$_3$ and Ca$_2$CuO$_3$ are characterized by very
anisotropic interaction strengths
\begin{equation}
J_{\parallel} \gg J_{\perp} \gg \mid J_{\perp,c}\mid ,
\label{jrelat}
\end{equation}
the anisotropy being about three orders of magnitude for each
inequality.  Instead of the spin-Peierls system CuGeO$_3$, in this
section for comparison we will consider the doped compound
GeCu$_{1-x}$Zn$_x$O$_3$ (x=0.034) \cite{hase96,poirier95} which shows
antiferromagnetic order. This is an example for an anisotropic
Heisenberg problem with weaker anisotropy than Sr$_2$CuO$_3$ and
Ca$_2$CuO$_3$. For simplicity, we will use for GeCu$_{1-x}$Zn$_x$O$_3$
the same parameters which were derived in the previous section for
CuGeO$_3$. We also neglect here the frustrated exchange $J_2$.

In the following we review several approaches for such anisotropic
systems where quantum and thermal fluctuations become important. We
will mention the usual spin wave approach in self-consistent random
phase approximation (RPA-SWA) where all directions are treated on an
equal and simple footing, and the coupled quantum spin chain
approach (CQSCA) which involves a sophisticated treatment of the
intra-chain direction and a mean-field treatment of the remaining
perpendicular inter-chain interactions.

\subsection{RPA spin wave theory}

The RPA-SWA yields simple analytical expressions for the N\'eel
temperature $T_N$ and for the magnetization $<S^z_A> = m_0$ at zero
temperature (see \cite{ami95}, \cite{majlis92} and references
therein). Both quantities can be derived from the expression
\begin{equation}
2 m(T)=\frac{1}{1+2\psi} \; , \quad
\psi=\frac{1}{N}\sum_{\vec{q}}\left(\frac{\omega_0}{\Omega(q)}\coth 
\left(\frac{\Omega(q)}{2T}\right) -1\right),
\label{RPASWA}
\end{equation}
where
\begin{equation}
\Omega(q)=\sqrt{ \omega_0^2 -\omega_1^2(q)} \; ,\quad 
\omega_1=4m(T) J_{\parallel}\left( \cos q_y +R \cos q_x +R_c 
\cos q_z \right),\\ 
\end{equation}
with $ \omega_0=4m(T)\left( J_{\parallel}+J_{\perp} +J_{\perp,c}
\right)$, $R=J_{\perp}/J_{\parallel}$ and $R_c
=J_{\perp,c}/J_{\parallel}$. We put $k_B=1$. The N\'eel temperature is
defined by the condition of vanishing magnetization which yields
\begin{equation}
2 T_N= J_{\parallel}/I(R,R_c),
\label{tn}
\end{equation}
where 
\begin{equation}
I(R,R_c)=\frac{1}{\pi^3}\int\!\!\int\!\!\int_0^\pi
\frac{dq_x dq_y dq_z}{R(1-\cos q_x) +R_c (1-\cos q_z)+ (1-\cos q_y)} \; . 
\label{tnint}
\end{equation}
Expanding $I(R,R_c)$ for $J_{\parallel} \gg J_{\perp} \gg J_{\perp,c}$
gives the approximate expression
\begin{equation}
I(R,R_c)
=\frac{0.66}{\sqrt{R}}
\left(1+0.24(\ln (R/R_c) \ -R)\right) \;, 
\label{eqi}
\end{equation}
which determines the N\'eel temperature together with (\ref{tn}).  The
zero temperature magnetization $m_0$ is in the same limit given by
\begin{equation}
m_0
=\frac{0.303}{1-0.386 \ln \left( R \right)} \; ,  
\label{mom}
\end{equation}
where the small parameter $R_c$ turned out to be irrelevant. Notice
that the RPA-description adopted reveals a vanishing magnetic moment
in the $R \rightarrow 0$ limit. Thus it differs from the ordinary
spin-wave theory which yields a diverging expression $m_0=\mid
0.5+(1/\pi)\ln R \mid$ in the weak inter-chain coupling limit.

Let us now check the above expressions using the estimates of the last
section and compare them with the experimental data. These data for
$T_N$ and the magnetic moment $\mu^{exp} = g_L m_0$ are given in the
Table. In the following we will adopt a typical cuprate Land\'e factor
$g_L \approx 2.1$ for Cu$^{+2}$ \cite{ami95}. Using the values
$J_{\parallel}$ and $J_{\perp}$ from the Table and
$J_{\perp,c}=10^{-4}$ meV, we find $T_N^{Sr}=38$ K, $T_N^{Ca}=75$ K,
$\mu^{Sr}=0.20 \mu_B$ and $\mu^{Ca} = 0.26 \mu_B$ for the Sr and Ca
compound, respectively. The ratio of the two experimental N\'eel
temperatures agrees approximately with the RPA-SWA prediction
\begin{equation}
T_N^{Ca}/T_N^{Sr} \approx 
\sqrt{J_{\parallel}^{Ca} J_{\perp}^{Ca}} / 
\sqrt{J_{\parallel}^{Sr} J_{\perp}^{Sr}} 
\approx 2
\label{ratio}
\end{equation}
where the logarithmic corrections in Eq.\ (13) can be neglected since
they are not very important for the above ratio. However, the absolute
values of $T_N$ within the RPA-SWA disagree  with the
experimental data.  In the case of the more isotropic
GeCu$_{1-x}$Zn$_x$O$_3$ we find $\mu^{Ge}=0.32 \mu_B$, i.e.\ a better
agreement. But also here, the magnetic moment is overestimated by the
RPA-SWA.  In this case that may be ascribed to the effect of the
frustrated second neighbor exchange.

For the Sr and Ca compounds one could try the opposite procedure using
the given experimental data (including $J_{\parallel}$) to determine
an ``empirical'' $J_{\perp}^{emp}$. Then one finds values for
$J_{\perp}^{emp}$ which are two (from $T_N$) or more than four (from
$m_0$) orders of magnitude lower than those estimated in the previous
section. This seems, therefore, to be unrealistic. Despite the fact
that it gives the correct limits for $m_0$ both for $R \to 0$ and in
the 2D isotropic case for $R \to 1$, the RPA-SWA seems to overestimate
$m_0$ for large anisotropy ($R\ll 1$) quite considerably. That points
to the necessity for an improved method. In the case of smaller
anisotropy (e.g.\ $R\sim 0.1$ like for GeCu$_{1-x}$Zn$_x$O$_3$), the
RPA-SWA seems to give more reliable results.

\subsection{Coupled quantum spin chain approach (CQSCA)}

Adopting Schultz's interchain RPA-expression (Eq.\ (7) of Ref.\
\cite{schulz96}), we replace $J_{\perp} \rightarrow 0.5(J_{\perp}
+J_{\perp,c})$ as would be suggested by our strongly ``orthorhombic''
parameter regime $J_{\parallel} \gg J_{\perp} \gg J_{\perp,c} $. Then
one arrives at
\begin{equation}
m_0= \gamma \sqrt{R}, 
\label{schmom}
\end{equation}
where the proportionality factor $\gamma$ is 0.72. (A similar factor
$2/\pi$=0.637, was obtained by Fukuyama {\it et
al.}\cite{fukuyama96}). The corresponding values for $\mu^{CSC}=g_L
m_0$ are listed in the Table.

Analogously, within these theories one expects $T_N \approx J_{\perp}$
\cite{affleck94}, in particular, the slightly modified implicit
expression for the transition temperature proposed by Schulz
\cite{schulz96} reads
\begin{equation}
T_N=\frac{2}{\pi}J_{\perp}\ln^{1/2}\left( \Lambda J_{\parallel}/T_N \right),
\label{tnsch}
\end{equation}
where $\Lambda \approx 5.8$.

From a principal point of view (Mermin-Wagner theorem), it is clear
that Eq.\ (\ref{tnsch}) overestimates $T_N$ because it does not depend
on $J_{\perp,c}$. However, if one assumes that its influence can be
described by logarithmic terms like in Eq.\
(\ref{eqi}) which then ensure a finite $T_N$, the relative changes 
might be quite small.

Like in the RPA-SWA, our estimated values for $J_{\perp}$ and the
experimental $J_{\parallel}$ lead to too large values for $T_N$ and
$m_0$. But now, using the experimental $m_0$ and $J_{\parallel}$ we
can determine from Eq.\ (16) an ``empirical'' $J_{\perp}^{emp}$ of the
CuO$_3$ chain compounds which is of a similar order of magnitude to
our estimates. The value of $J_{\perp}^{emp}$ is smaller by a factor
between 2 and 3 (Sr$_2$CuO$_3$, GeCu$_{1-x}$Zn$_x$O$_3$), or 6
(Ca$_2$CuO$_3$) compared to the theoretically estimated values (see
Table). The N\'eel temperature can also be used to determine
$J_{\perp}^{emp}$ which gives similar values showing the internal
consistency of the CQSCA. But one should keep in mind that Eq.\ (18)
does not fulfill the Mermin-Wagner theorem. In that respect we note
here an alternative approach to the strongly anisotropic Heisenberg
model (Castro Neto and Hohn \cite{castro96}) in which the N\'eel
temperature was found to depend linearly on $J_{\perp,c}$. Naturally,
the elucidation of the correct description as to how this smallest
interaction parameter does affect the finite temperature properties
remains a challenging problem. Without its generally accepted solution
it makes no sense to discuss the absolute values of the
N\'eel-temperature beyond an order of magnitude accuracy.

One possible explanation for the reduction of $J_{\perp}^{emp}$
in comparison with our estimated $J_{\perp}$ could be the proximity of
a spin-Peierls state.  Phase fluctuation effects beyond the mean-field
inter-chain approach used in deriving Eq.\ (\ref{schmom}) can then
become quite important. Following the renormalization group approach
of Wang \cite{wang96} for a plane of weakly interacting chains at
$T=0$, one finds a strongly renormalized magnetization which can be
traced back to a renormalized exchange integral. If that is true, one
should expect Ca$_2$CuO$_3$ to be much closer to the SP phase
transition point than Sr$_2$CuO$_3$. Furthermore for small exchange
integrals compared with the phonon frequency ($\sim 10$ to 20 meV),
phonon exchange gives rise to a quasi-instantaneous interaction
between localized spins, leading to a renormalization $J \rightarrow
J_{eff} < J$ \cite{schulz87}.

Another possible origin for the difference between $J_{\perp}$ and
$J_{\perp}^{emp}$ might be our simple procedure to estimate
$J_{\perp}$ based on the extended Hubbard model. It was already
mentioned that such a procedure has the tendency to overestimate the
exchange integrals which becomes already apparent for $J_{\parallel}$.
Last not least, one should keep in mind an uncertainty of the
band-structure methods with respect to transfer integrals as small as
in the considered case. The replacement of the full potential in
the region in between the chains by empty spheres as explained in
Sec.\ II might effect the transverse tails of the Wannier-functions
which determine the value of the transfer integral $t_{\perp}$. A
similar effect is expected if the exchange and correlational potential
is supplemented with gradient terms reflecting the strong change in
the electron density moving from one chain to its neighbors.  Anyhow,
one would expect roughly the same accuracy for the Sr and the Ca
compound. In this context the much stronger deviation of the magnetic
moment of the Ca compound might be related to somewhat reduced
accuracy of Eq.\ (16) in less anisotropic cases.

\section{Summary} 

Band-structure calculations for Sr$_2$CuO$_3$ and \cco\ show in
addition to the expected large dispersion along the chain direction
also a remarkable dispersion in the crystallographic {\bf
a}-direction, i.e.\ perpendicular to the plane containing the
corner-shared CuO$_4$ plaquettes which form the CuO$_3$ chains. The
corresponding inter-chain transfer gives rise to antiferromagnetic
exchange integrals $J_{\perp}$ in the meV range. Together with a small
dipolar exchange in the third direction $J_{\perp,c}$ it explains the
antiferromagnetic order in terms of an anisotropic Heisenberg model.
The larger value of $J_{\perp}$ for Ca$_2$CuO$_3$ corresponds with the
larger N\'eel temperature and the larger magnetic moment
\cite{kojima97} in comparison with Sr$_2$CuO$_3$. However, our rough
estimation of exchange integrals, based on the one band assumption,
seems to overestimate the difference between both
substances. CuGeO$_3$ is different from these two substances by a much
smaller exchange in chain direction and a comparable large frustration
parameter which suppresses the antiferromagnetic state and stabilizes
the spin Peierls state.  The smaller anisotropy becomes apparent in
GeCu$_{1-x}$Zn$_x$O$_3$ where a N\'eel state was found with
significantly larger magnetic moments than in Sr$_2$CuO$_3$ or
Ca$_2$CuO$_3$.

The copper oxygen chains of the three compounds under consideration
can be described within an extended Hubbard model supplemented by
ferromagnetic contributions to the nearest neighbor exchange
integral. For the Ca and Sr compounds excitonic effects in the limit
$q\rightarrow 0$ are expected to be weak due to small inter-site
Coulomb interaction $V_1$. But near the zone boundary $q \approx
\pi/b$ strong excitonic effects are expected in the framework of the
theory developed recently by Stefan and Penc \cite{stephan96}.  If our
proposed parameterization is correct, for Ca$_2$CuO$_3$ ($\approx$3.5
eV) and CuGeO$_3$ ($\approx$ 4.3 eV) these excitonic peaks should be
observed at higher energies than for Sr$_2$CuO$_3$ (3.15 eV).

The LDA band structure calculations yield useful insights into
important material dependent parameters as inter-chain electron
transfer and tendencies of the crystal field (Madelung ) potential,
varying from one substance to the other, albeit that the estimate of
the on-site and inter-site Coulomb interaction requires more
sophisticated methods such as LDA-calculations with local constraint.

According to our findings, Ca$_2$CuO$_3$ is expected to be a typical
1D charge transfer insulator analogously to the 2D model system
Sr$_2$CuO$_2$Cl$_2$ although the inter-chain interaction is
intermediate between the strongly correlated CuGeO$_3$ and the most 1D
Sr$_2$CuO$_3$.  Possibly, the latter system is the weakest correlated
one of the three. The compounds Sr$_2$CuO$_3$ and Ca$_2$CuO$_3$ offer
also the opportunity to study in detail the effect of the inter-chain
interaction provided it can be changed in a controlled way. Indeed,
the study of the magnetic properties of the alloy system
Sr$_{2-x}$Ca$_x$CuO$_3$ gives an interesting possibility to change
continuously the magnitude of the inter-chain coupling. This is also
interesting from the theoretical point of view since it gives a
possibility to check in more detail sophisticated theories for weakly
coupled quantum spin chains.

\acknowledgements 
Discussions with Profs.~J.\ Fink, P.\ Fulde, D.\ Hone, H.\ Schulz, D.\
Johnston, and T.\ Van Oosten are acknowledged.  Special thanks to Dr.\
M.\ S.\ Golden for discussions and a critical reading of the manuscript
and to R.\ Neudert providing us with EELS data for Sr$_2$CuO$_3$ prior
to publication.  One of us (J.\ M.)  thanks the Max-Planck-Institut
``Komplexe Systeme'' Dresden, for hospitality during which part of the
present work was performed.  Finally, the Deutsche
Forschungsgemeinschaft is also acknowledged for financial support
(S.-L.\ D. and J.\ M.)  .

\newpage
\table{\ TAB.\ Model parameters for Sr$_2$CuO$_3$, Ca$_2$CuO$_3$ and
CuGeO$_3$.  The LDA-LCAO derived tight binding parameters in the first
group of rows are explained in Sec.\ II.  The LDA-numbers for
CuGeO$_3$ are estimated from Figs.\ 2 of Ref.\ \cite{mattheiss94} and
Ref.\ \cite{popovic95}.  The second group contains experimental values
(in the case of several date we prefer the underlined) which were used
in addition to the band-structure information to estimate the
corresponding parameters of the extended Hubbard model as well as the
exchange integrals of the anisotropic Heisenberg-model (third group of
rows). They are derived and discussed in Sec.\ III.  The experimental
magnetic moment $\mu^{exp}$ (given together with $T_N$ in group IV)
may be compared with $\mu^{CSC}$ derived from Eq.\ (19) using the
experimental data for the in-chain exchange integrals $J_{\parallel}$
and our estimation of $J_{\perp}$. Vice versa, the experimental
$\mu^{exp}$ determines via the same Eq.\ (19) the empirical
inter-chain exchange integrals $J^{emp}_{\perp}$.  }\\

\newpage

\begin{tabular}{c|c||c|c|c}
group & quantity& Sr$_2$CuO$_3$ & Ca$_2$CuO$_3$ & CuGeO$_3$ \\
      &          &               &     & (GeCu$_{1-x}$Zn$_x$O$_3$) \\[12pt]
\hline
\hline
I & $t_{1,LDA}/$meV & 550 & 520 &250\\[12pt]
  & $t_{2,LDA}/$meV & 100& 100& 81\\[12pt]
  & $t_{\perp}/$meV &20 to 30 & 50 to 65  & 25 to 33\\[12pt]
\hline
II & $E_g/$eV        & (1.8 to 1.9)$^{\rm a,b}$  & (2.1)$^{\rm c}$
     & (1.25)$^{\rm d}$, 
     ($\underline{3.7}$)$^{\rm e}$ \\[12pt]
  & $J/$meV ($=J_{\parallel}/$ meV) &(140)$^{\rm f}$,
    ($\underline{190}$)$^{\rm g}$, (260)$^{\rm h}$  
    & ($\underline{160}$)$^{\rm i}$ (254)$^{\rm h}$
    & ($\underline{11}\pm 1$)$^{\rm j}$,    (22 )$^{\rm k}$   \\[12pt] 
\hline
III &  $t_1/$meV & 410 & 419 &  187\\[12pt]
  &  $t_2/$meV & 100 & 100 &  90\\[12pt]
  &  $U_{eff}/$eV  & (3.15)$^{\rm b}$ & 3.5 &  4.34 \
        (4.2)$^{e}$\\[12pt] 
  & $V_1/$eV & 0.21 & 0.16 & 0.1\\[12pt]
  & $\mid K \mid$/meV & 11 & 30 & 19 \\[12pt]
  & $J_{\perp}/$meV&0.5 to 1.1&2.9 to 4.3&0.6 to 1, (1.1)$^l$ \\[12pt]
\hline
IV & $T_N/$K & (5$)^l$ & (8...10)$^l$ & (4.5)$^m$ \\[12pt]
  &$\mu^{exp}/\mu_B$ & (0.06$\pm 0.01$)$^{l}$ & (0.09$\pm 0.01$)$^l$ &
       (0.23)$^j$ \\[12pt] 
\hline
V  & $\mu^{CSC}/\mu_B$ & 0.08 to 0.11 & 0.19 to 0.24 & 0.35 to 0.45  \\[12pt]
 & $J_{\perp}^{emp}/$meV&0.3$\pm$ 0.1 &0.6$\pm 0.1$& 0.27 \\[12pt]
\end{tabular}

\noindent
$^{\rm a}$\ Raman resonance, Ref.\ \cite{misochko96}\\
$^{\rm b}$\ EELS, Ref.\ \cite{neudert97}\\
$^{\rm c}$\ opt.\ absorption, Ref.\ \cite{tokura90}\\
$^{\rm d}$\ XPS, Ref.\ \cite{terasaki95}\\
$^{\rm e}$\ XPS, Ref.\ \cite{parmigiani96}\\
$^{\rm f}$\ magn.\ suscept., Ref.\ \cite{ami95,eggert96}\\
$^{\rm g}$\ magn.\ suscept., Refs.\ \cite{motoyama96}\\
$^{\rm h}$\ midinfrared, Ref.\ \cite{suzuura96}\\
$^{\rm i}$\ theory, Ref.\ \cite{drechsler96}\\
$^{\rm j}$\ INS, Raman, Refs.\ \cite{regnault96,gros96,hase96,muellerfledder96}\\
$^{\rm k}$\ Raman, Ref.\ \cite{kuroe96}\\
$^{\rm l}$\ $\mu$SR, Ref.\ \cite{kojima97}\\
$^{\rm m}$\ INS, Ref.\ \cite{nishi95}\\

\newpage

\figure
{FIG.\ 1\ Crystal structure of Sr$_2$CuO$_3$
\label{fig1}}
\figure
{FIG.\ 2\ LCAO energy-bands near the Fermi level for
Sr$_2$CuO$_3$ along high-symmetry Brillouin-zone directions
within the ($k_x,k_y$) plane. The momenta are given in units of
($\pi/a,\pi/b$).
Strong 
dispersion can be seen along  (0,0)--(0,1) ({\bf b}-direction, 
 parallel to the CuO$_3$ chains)  whereas a small, but 
non-negligible dispersion in  the perpendicular
{\bf a}-direction can be seen. 
\label{fig2}}
\figure
{FIG.\ 3\ 
Density of states $N(E)$ for Sr$_2$CuO$_3$. The insert shows the
partial Cu $3d$ and $4s$ density of states of the nearly
one-dimensional band crossing the Fermi level.
\label{den}}
\figure
{FIG.\ 4\
Dispersion of the nearly one-dimensional band. The insert is for fixed
$k_y$ = $k_{yFermi}$ = $\pi/2b$.
 \label{fig4}}  
\figure
{FIG.\ 5\
Dependence of the transfer integral $t_1$ (upper panel) and the on-site
Coulomb interaction $U_{eff}$ (lower panel) of the Hubbard model according to its
Bethe ansatz-solution  {\it vs}. 
inchain superexchange integral $J_{1}^{AF}$ (lower panel) in the strong coupling 
limit
for typical values of the optical gap $E_g$. The experimental 
values for $J_{1}^{AF}$ are depicted by arrows. They are determined
from the total exchange integral $J$ adopting ferromagnetic and second
neighbor contributions discussed in the text.
\label{plot}}

\begin{references}
\bibitem[21]{remark4}
Adopting the applicability of the RPA-spin wave theory and the value 
of the 
weak dipolar interchain interaction of about 0.01 meV evaluated 
in \cite{ami95} one would arrive at $T_N=6.2$ K, i.e. accidentily 
close to 
the experimental value, in sharp contrast to the stated $T_N \approx
$0.028 K, only.
Using the more realistic interchain value of about 1 meV obtained below, one
arrives formally at $T_N \approx 29$ K. For a critical discussion of that 
approach see Sec.\ IV.
\bibitem[50]{remark2}
The attribution of sharp peaks near 1.25 eV, 2.9 eV 
and 3.66 eV observed in Im$\varepsilon (\omega)$ to the $dp$ 
charge transfer gap which is of interest here, to $dd$-transitions, and to  
Ge-states related interband transition is controverse.
\end{references}
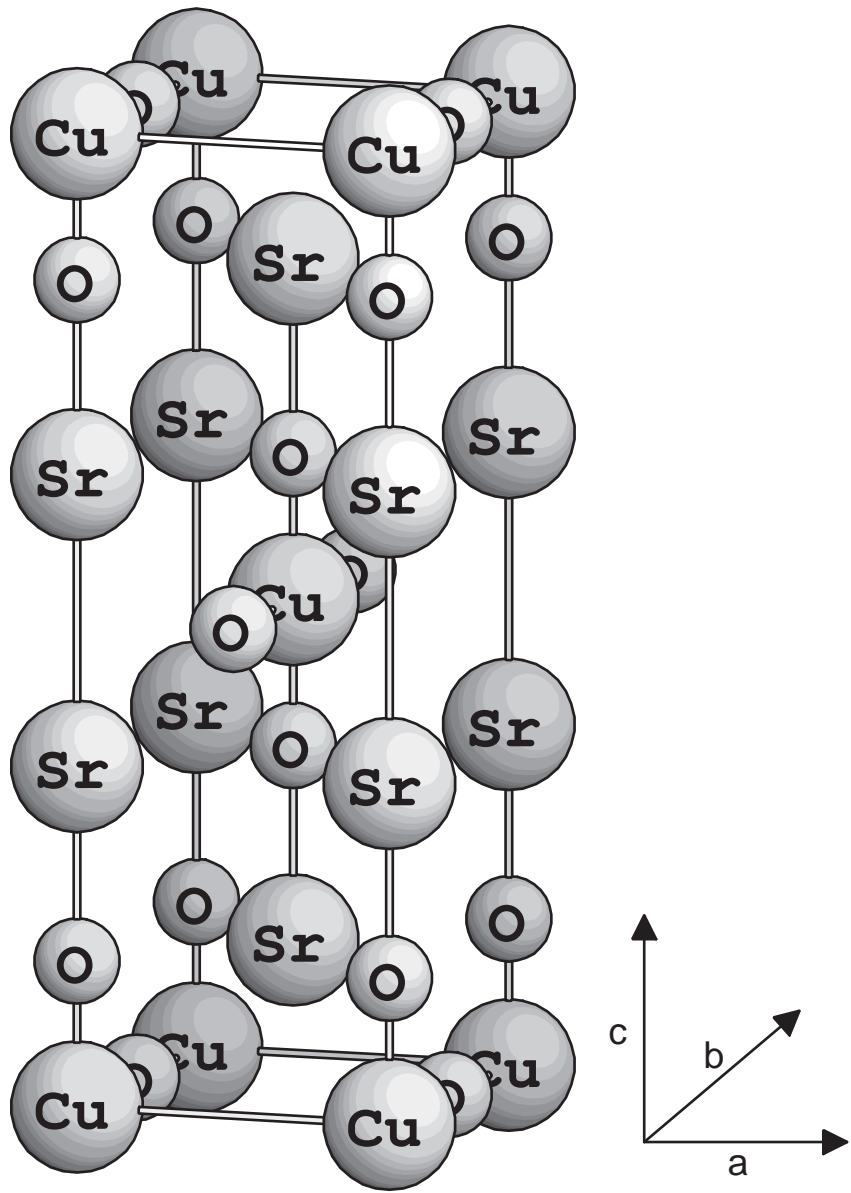
\end{document}